\documentclass[11pt,twoside]{article}

\usepackage{prcsa2008}
\usepackage{epsf}
\usepackage{psfig}
\usepackage{lscape}

\usepackage{graphicx}

\begin{document}
\title{SkyMapper and the Southern Sky Survey: a valuable resource for stellar astrophysics}  
\author{Simon Murphy, Stefan Keller, Brian Schmidt, Patrick Tisserand, Michael Bessell, Paul Francis and Gary Da Costa}   
\affil{Research School of Astronomy and Astrophysics\\ Australian National University\\ Cotter Road, Weston Creek, ACT 2611, Australia}    

\begin{abstract} 
The Australian National University's SkyMapper telescope is amongst the first of a new generation of dedicated wide-field survey telescopes. Featuring a 5.7 deg$^{2}$ field-of-view Cassegrain imager and 268 Mega-pixel CCD array, its primary goal will be to undertake the Southern Sky Survey: a six color ($uvgriz$), six-epoch digital record of the entire southern sky. The survey will provide photometry for objects between 8th and 23rd magnitude with a global photometric accuracy of 0.03 magnitudes and astrometry to 50 mas. In this contribution we introduce the SkyMapper facility, the survey data products and outline a variety of case-studies in stellar astrophysics for which SkyMapper will have high impact.
\end{abstract}

\section{Introduction}

It is now possible for CCD mosaic imagers on small telescopes to achieve areal coverage that was once solely the domain of photographic Schmidt plates. Digital surveys offer better photometric and astrometric precision and calibrations, in part due to the excellent linearity and uniformity of modern CCD detectors. Several groups around the world are now actively pursuing digital, multi-colour surveys of the sky. The SkyMapper telescope and Southern Sky Survey fill an urgent demand for high \emph{etendue} multi-colour optical survey work in the southern hemisphere that will be unrivaled until the commissioning of the Large Synoptic Survey Telescope (LSST) in Chile in 2015.

\section{The SkyMapper Telescope}

The SkyMapper telescope is being constructed  
by the Australian National University's Research School of Astronomy and 
Astrophysics, in conjunction with Electro Optic Systems of Canberra, Australia. Situated at Siding Spring Observatory in New South Wales, Australia (latitude 31.3\deg~S, altitude 1169 m), the telescope is a modified Cassegrain design with a 1.33 m primary aperture and  0.69 m secondary. This optical design facilitates a large 5.7 deg$^{2}$ field-of-view with a working focal ratio of f/4.78.

The focal plane is serviced by a 268 Mega-pixel mosaic imager of 32 2k~$\times$~4k E2V CCDs.  Each CCD has near-perfect cosmetics and excellent quantum efficiency over the 350--950 nm wavelength range. The 0.5 arcsec pixels (15$\mu$m) are well matched to the 1.1 arcsec median seeing at the site. The entire facility and data reduction pipeline will operate in an automated manner with minimum operational support required. The telescope is currently undergoing final integration and will see first light in late 2008. Comprehensive  information about all aspects of the SkyMapper programme can be found in \cite{SM_Keller07}.

\section{Southern Sky Survey}

The primary role of SkyMapper will be to undertake the Southern Sky Survey. This six-colour, six-epoch survey will cover the entire 2$\pi$ steradian of the southern sky, reaching depths of $g=23$ mag at a detection limit of 5$\sigma$. For stars brighter than $g=18$ we require a photometric precision of 0.03 mag globally and astrometry to better than 50~mas. This equates to proper motions better than 4~mas~yr$^{-1}$ over the five year baseline of the survey. 

The survey's six epochs are designed to capture variablity on timescales of hours (e.g.~RR Lyraes, asteroids), days, weeks, months (supernovae, long-period variables) and years (QSOs, parallax, proper motions). Exposure times are 110~s, a balance between survey depth and the need for a timely survey completion. The expected magnitude limits in typical 1.5 arcsec seeing for a 5$\sigma$ detection after a single 110 s exposure and after coadding the full six epochs are shown in Table~\ref{SM_tab:limits}. In all bands we achieve limits some $\sim$0.5 mag deeper than the Sloan Digital Sky Survey.

\begin{table}[tb!]
   \centering
     \begin{tabular}{lcccccc}
       \tableline
       \noalign{\smallskip} 
	 & $u$ & $v$ & $g$ & $r$ & $i$ &$z$ \\
	 \noalign{\smallskip} 
      \tableline
       \noalign{\smallskip} 
      1 epoch & 21.5 & 21.3 & 21.9 & 21.6 & 21.0 & 20.6 \\
      6 epochs & 22.3 & 22.7 & 22.9 & 22.6 & 22.0 & 21.5\\
 \noalign{\smallskip} 
          \tableline
   \end{tabular}
         \caption{Expected Southern Sky Survey magnitude limits (at 5$\sigma$ signal-to-noise) for 110 s exposures. Magnitudes are in the AB system.}
    \label{SM_tab:limits}
\end{table}

\subsection{Science Goals}

Although the range of possible science results from the Southern Sky Survey cannot be fully anticipated or described we have identified a set of science goals which SkyMapper will have high impact in addressing. These include finding the most extremely metal-poor stars in the Galaxy, determining the shape and extent of the Galactic dark matter halo, the distribution of solar system objects beyond Neptune, the discovery of dozens of high-redshift ($z>5$) quasars and revealing the star-formation history of the youngest stars in the solar neighbourhood. Further information may be found in \citet{SM_Keller07}.

The Southern Sky Survey will occupy approximately 75\% of available telescope time. The remainder is available for non-survey science in collaboration with RSAA researchers. This includes a capacity for targets of opportunity. 

\section{A Filter Set for Stellar Astrophysics}

\begin{figure}[tb!] 
   \centering
   \includegraphics[width=0.6\linewidth]{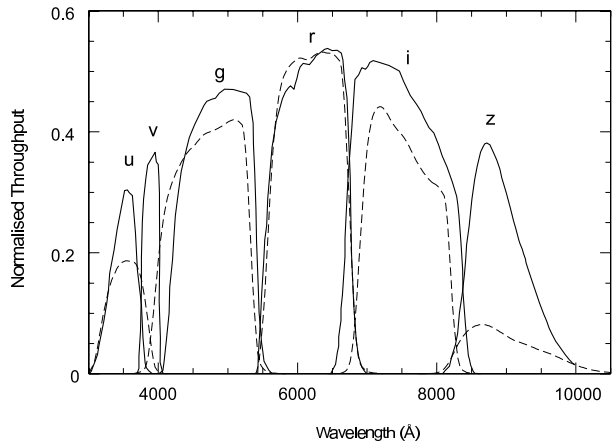} 
   \caption{Throughput of the SkyMapper filter set, excluding atmospheric absorption. Shown for comparison are the SDSS bandpasses (dashed lines). Note the higher $u$ and $z$ throughput of the SkyMapper filters.}
   \label{SM_fig:filters}
\end{figure}

Many of the science goals outlined above rely on a clean identification of stellar populations. It was therefore important that we chose a filter set that offers optimal diagnostic power for the important stellar parameters of effective temperature, surface gravity and metallicity. Through comparison with model atmospheres we have converged at the filter set shown in Figure~\ref{SM_fig:filters}. The filters are broadly based on the $griz$ bands used by the Sloan Digital Sky Survey, with an additional Str\"omgren-like $u$ filter and intermediate-band $v$ filter (similar to the DDO-38) bracketing the Balmer Jump feature at 3646 \AA. Compared to the SDSS system the SkyMapper filters offer twice the throughput in $u$ and three times the throughput in $z$, albeit with our smaller aperture. The addition of the $u$ and $v$ filters  means we are able to break the degeneracy between stellar surface gravity and metallicity, as illustrated in the following examples.

\subsection{Case Study I: Blue Horizontal Branch Stars}

\begin{figure}[tb!] 
   \centering
   \includegraphics[width=0.6\linewidth]{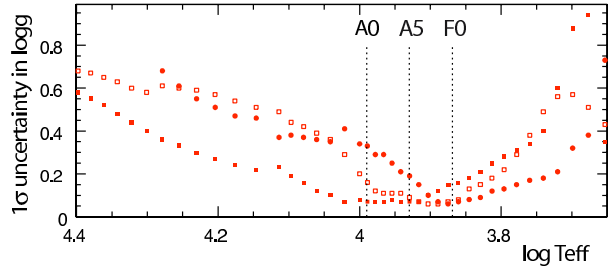} 
   \caption{Uncertainty in surface gravity derived under photometric uncertainties of 0.03 mag in the SkyMapper filter set. Symbols are $\log g = 4.5$ (solid squares), $\log g = 3.5$ (open squares) and $\log g = 2.0$ (circles).}
   \label{SM_fig:logg_unc}
\end{figure}

Figure~\ref{SM_fig:logg_unc} shows the expected uncertainty in the derived stellar surface gravity as 
a function of temperature, assuming typical photometric 
uncertainties of 0.03~mag per filter. For A-type stars we expect 
to determine $\log g$ to $\sim$10\%. The sensitivity to gravity 
arises primarily from the $u-v$ colour, which measures the Balmer Jump and the effect 
of H$^{-}$ opacity, both of which increase with surface gravity. It is in this 
temperature range that we find Blue Horizontal Branch stars (BHBs). The 
characteristic absolute magnitude of BHBs allows them to be used as standard candles for the Galactic 
halo \citep[e.g.][]{SM_Newberg07}.

\begin{figure}[tb!] 
   \centering
   \includegraphics[width=0.6\linewidth]{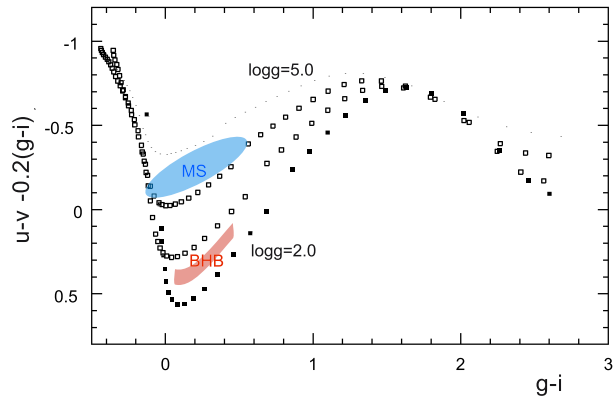} 
   \caption{$u-v$ vs. $g-i$ for stars of solar metallicity and a range of surface gravities. Blue Horizontal Branch stars are well separated from main sequence and blue straggler stars.}
   \label{SM_fig:bhb_ms}
\end{figure}

A line of sight through the halo inevitably contains a mixture of local 
main-sequence A-type and blue straggler stars. However, as shown in Figure~\ref{SM_fig:bhb_ms} the SkyMapper filter set enables us to clearly distinguish the BHBs of 
interest on the basis of their lower surface gravity. Simulations show that 
we will be able to derive a sample of BHBs out to 130 kpc with less than 5\% 
contamination. 

\subsection{Case Study II: Extremely Metal-Poor Stars}

\begin{figure}[tb!] 
   \centering
   \includegraphics[width=0.55\linewidth]{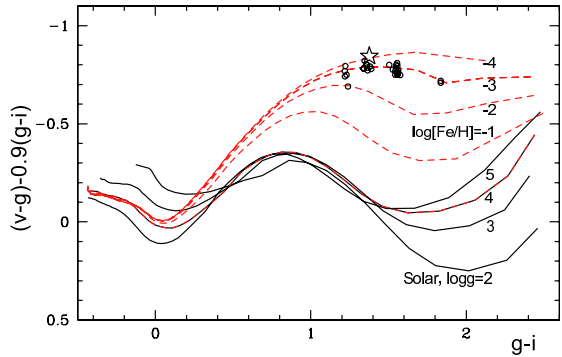} 
   \caption{$v-g$ vs. $g-i$ for stars of solar metallicity and a range of surface gravities (solid lines) and for $\log g = 4$ and a range of subsolar metallicities (dashed lines). Overlaid are the computed colours of HE1327-2326 (open star), the most metal-poor star currently known \citep{SM_Frebel05} and the metal-poor sample of \citet{SM_Cayrel04} (open circles).}
   \label{SM_fig:feh_tracks}
\end{figure}

An analogous plot to Figure~\ref{SM_fig:logg_unc} can be made for the uncertainty in derived stellar metallicity from SkyMapper photometry. Such an analysis shows we are best able to determine [Fe/H] at high (solar) metallicities, with the uncertainty rising to $\pm0.7$ dex at [Fe/H]~=~-4. At these extremely low metallicities the number density of stars drops by a factor of ten for every dex of metallicity. At the temperature range inhabited by evolved Extremely Metal-Poor stars the $v-g$ colour is insensitive to $\log g$ but very sensitive to metallicity. Figure~\ref{SM_fig:feh_tracks} demonstrates that SkyMapper photometry is more than able to separate the bulk of halo stars at [Fe/H]~$>$~-2 from the rare and interesting EMPs. From simulations we expect to discover $\sim$1000 stars with [Fe/H]$<-4$ and $\sim$100 having [Fe/H]$<-5$ at magnitudes brighter than $g=18$ over the entire survey area.

\subsection{Case Study III: H$\alpha$ Emission Line Objects}

We have seen throughout this conference that Balmer emission is a characteristic of many interesting stellar systems. Major groups of emission-line stars include evolved massive stars (including Wolf-Rayet and Be stars), post-asymptotic giant branch stars, pre-main sequence stars, active stars, novae and binary systems (e.g. symbiotic, cataclysmic variables). Although not part of the standard SkyMapper filter set, non-survey narrowband H$\alpha$ imaging will also be possible thanks to the addition of the large-format H$\alpha$ filter used previously in the SuperCosmos H$\alpha$ Survey \citep{SM_Parker98}. When combined with an $r$-band measurement of the surrounding continuum, systems of interest can be easily separated from the fiducial stellar locus \citep[see for example Figure~1 in][and Figure 2 in Murphy \& Bessell, these proceedings]{SM_Corradi08}.

\section{Data Products}

To avoid costly recalibrations to previously released data the survey data products will only be released after extensive quality control. The first SkyMapper data product released will be the Five Second Survey -- a three-epoch survey of 8th to 16th magnitude stars in all filters under photometric conditions. This is expected to be completed by mid-2010. Two Main Survey data releases are anticipated, the first after three observations in each filter are completed for all fields, and the final release when all six epochs are available (reaching a depth of $g\sim23$). These are expected at one and five years after observations begin, respectively. Data in the form of object catalogues and flat-fielded, WCS calibrated images will be provided to the community via the Virtual Observatory initiative and the survey website (\texttt{http://www.mso.anu.edu.au/skymapper}).


\begin{thebibliography}{}
\bibitem[Cayrel et 
al.(2004)]{SM_Cayrel04} Cayrel, R., et al.\ 2004, \aap, 416, 1117 
\bibitem[Corradi et 
al.(2008)]{SM_Corradi08} Corradi, R.~L.~M., et al.\ 2008, \aap, 480, 409 
\bibitem[Frebel et al.(2005)]{SM_Frebel05} Frebel, A., et al.\ 
2005, \nat, 434, 871 
\bibitem[Keller et al.(2007)]{SM_Keller07}
Keller, S.~C., et al.\ 
2007, PASA, 24, 1 
\bibitem[Parker 
\& Bland-Hawthorn(1998)]{SM_Parker98} Parker, Q.~A., \& Bland-Hawthorn, J.\ 1998, PASA 15, 33 
\bibitem[Newberg et al.(2007)]{SM_Newberg07} Newberg, H.~J., et al.\ 2007, \apj, 668, 221 
\end{thebibliography}
\end{document}